\documentclass{article}
\usepackage{graphicx}
\usepackage{amsmath}
\usepackage{amsfonts}
\usepackage{amssymb}

\begin{document}

\title{Lower energy bounds for quantum lattice Hamiltonians}
\author{Dean Lee$^{1}$and Nathan Salwen$^{2}$\\{\small Department of Physics, North Carolina State University, Raleigh, NC 27695}\\$^{1}${\small dean\_lee@ncsu.edu, }$^{2}${\small nathan\_salwen@ncsu.edu}}
\maketitle
\begin{abstract}
We derive general lower energy bounds for the ground state energy of any
translationally invariant quantum lattice Hamiltonian. \ The bounds are given
by the ground state energy of renormalized Hamiltonians on finite clusters.
\end{abstract}

\section{Introduction}

When computing the low energy eigenstates for some quantum Hamiltonian, $H$,
one usually extracts an energy estimate by calculating the expectation of $H$
for some approximate wavefunction. In the case of the ground state this
estimate also provides an upper bound on the energy. \ The calculation is
usually repeated with systematically improved wavefunction approximations and
the final result is determined by extrapolation. \ While in principle this
procedure should work, in practice computational difficulties can block
improvement beyond some point. \ One common culprit is the fermion sign
problem. \ Although the sign problem can be delayed with some success, as
shown in \cite{lee}, in the end it will eventually lead to exponential scaling
of computational effort. \ In such cases it is quite useful to have some idea
of how far a given estimate is from the exact answer.

Since in most cases the result in hand provides an upper bound to the ground
state energy, what is needed is a reasonably accurate lower bound. \ In this
letter we state and derive a general lower bound for the ground state energy
of any translationally invariant lattice Hamiltonian. \ Our result is very
general and applies to any number of dimensions, size of lattice, type of
particle, or form of local interactions. \ In particular this includes
single-site terms, nearest and next-nearest neighbor hoppings, and gauge
plaquettes. \ The only restriction is that the interactions are localized and
do not extend throughout the entire lattice. \ The lower bound is stated in
terms of the ground state energy for a renormalized Hamiltonian defined on a
cluster subset. \ As we find in the example described below, the error appears
to scale as the ratio of the boundary size to the bulk size of the cluster.
\ Since the lower bound is defined on clusters, it is most convenient for
position-space computational schemes which already partition the lattice into
clusters. \ This includes density matrix renormalization group methods
\cite{white} and Monte Carlo simulations using Hamiltonian partitioning
\cite{hirsch}$.$

Our letter is organized as follows. \ We begin by proving a general inequality
involving two Hamiltonians which are equivalent within a projection subspace.
\ We then apply the general inequality to a simple but illustrative example
involving a one-dimensional periodic chain Hamiltonian and a non-periodic
chain Hamiltonian. \ We then state and prove our main result, the lower energy
bound. \ To conclude we check the lower bound numerically for our
one-dimensional example and note the scaling of the error with cluster size.

\textit{After posting this preprint to the archive, it has come to our
attention that the lower bound techniques presented here have already been
discussed in the literature \cite{anderson}-\cite{valenti}.}

\section{General inequality}

We prove a simple but useful result from which the lower energy bound follows
as a special case. \ Let $H$ be a self-adjoint Hamiltonian. \ Let $P$ be a
projection operator that projects onto a subspace $V_{P}$. \ We assume that
$P$ commutes with $H$. \ Let $H^{\prime}$ be any self-adjoint operator whose
spectrum is bounded below and satisfies%
\begin{equation}
PHP=PH^{\prime}P.
\end{equation}
Let $E_{0}^{\prime}$ be the ground state energy of $H^{\prime}$. \ For any
vector $v$ we find
\begin{equation}
\left\langle Pv,HPv\right\rangle =\left\langle v,PHPv\right\rangle
=\left\langle v,PH^{\prime}Pv\right\rangle =\left\langle Pv,H^{\prime
}Pv\right\rangle .
\end{equation}
Since
\begin{equation}
\left\langle Pv,H^{\prime}Pv\right\rangle \geq E_{0}^{\prime}\left\langle
Pv,Pv\right\rangle ,
\end{equation}
we conclude that
\begin{equation}
\left\langle Pv,HPv\right\rangle \geq E_{0}^{\prime}\left\langle
Pv,Pv\right\rangle . \label{bound}%
\end{equation}
Therefore the spectrum of $H$ restricted to $V_{P}$ is bounded below by
$E_{0}^{\prime}$.

\section{Application to lattice Hamiltonians}

We now apply the general result (\ref{bound}) to a translationally invariant
lattice Hamiltonian. \ For clarity we consider a simple but sufficiently
general example. \ Our example is a length $N$ one-dimensional periodic chain
with one type of fermion. \ We define $O_{1}$ as a hopping term to and from
the origin
\begin{equation}
O_{1}=a_{1}^{\dagger}a_{0}+a_{0}^{\dagger}a_{1}\text{,}%
\end{equation}
and $O_{2}$ as the number operator at the origin,
\begin{equation}
O_{2}=a_{0}^{\dagger}a_{0}\text{.}%
\end{equation}
Our Hamiltonian describes a free one-dimensional fermion,
\begin{equation}
H=c_{1}\sum_{n=0,...,N-1}\left[  a_{n+1}^{\dagger}a_{n}+a_{n}^{\dagger}%
a_{n+1}\right]  +c_{2}\sum_{n=0,...,N-1}a_{n}^{\dagger}a_{n}. \label{hamilton}%
\end{equation}

Let us now take a four-site non-periodic chain and the\ renormalized
Hamiltonian $H^{\prime}$,
\begin{equation}
H^{\prime}=\tfrac{c_{1}}{3}\sum_{n=0,1,2}T_{n}O_{1}+\tfrac{c_{2}}{4}%
\sum_{n=0,1,2,3}T_{n}O_{2}. \label{three}%
\end{equation}
The reason for the renormalization of coefficients will be apparent shortly.
\ We define $E_{0}^{\prime}$ as the ground state energy of $H^{\prime}$ and
take $P$ as the projection operator
\begin{equation}
P=\tfrac{1}{N}\sum_{n=0,...,N-1}T_{n}\text{.}%
\end{equation}
We note that
\begin{align}
PH^{\prime}P  &  =\tfrac{c_{1}}{3}\cdot3\cdot PO_{1}P+\tfrac{c_{2}}{4}%
\cdot4\cdot PO_{2}P\\
&  =\tfrac{1}{N}PHP\text{.}\nonumber
\end{align}
From (\ref{bound}) we conclude\ that the spectrum of $H$ in the
translationally invariant subspace is bounded below by $N\cdot E_{0}^{\prime}$.

\section{Lower energy bound}

We now state the general result, a lower energy bound for finite periodic
lattices of any size or number of dimensions. \ Let $T_{\overrightarrow{r}}$
be the operator corresponding with translation by the vector $\overrightarrow
{r}$. \ We observe that a local operator such as $a_{\overrightarrow{s}}$
transforms as $T_{\overrightarrow{r}}a_{\overrightarrow{s}}T_{-\overrightarrow
{r}}=a_{\overrightarrow{r}+\overrightarrow{s}}$. \ Let $G$ be any subgroup of
lattice translations $T_{\overrightarrow{r}}$ and $\left|  G\right|  $ be the
number of elements of $G$. Let $H$ be a lattice Hamiltonian defined by%
\begin{equation}
H=\sum_{T_{\overrightarrow{r}}\in G}\left[  \sum_{j}c_{j}T_{\overrightarrow
{r}}O_{j}T_{-\overrightarrow{r}}\right]  , \label{h}%
\end{equation}
where the $c_{j}$'s are coefficients and $O_{j}$'s are general quantum
operators. \ Let $H^{\prime}$ be a modified Hamiltonian defined on a subset of
lattice points,
\begin{equation}
H^{\prime}=\sum_{j}\left[  \tfrac{c_{j}}{n_{j}}\sum_{k=1,...,n_{j}%
}T_{\overrightarrow{r}_{j,k}}O_{j}T_{-\overrightarrow{r}_{j,k}}\right]  .
\label{renorm}%
\end{equation}
The $n_{j}$'s are arbitrary positive integers and each $\overrightarrow
{r}_{j,k}$ is a lattice vector (not necessarily distinct). \ Let
$E_{0}^{\prime}$ be the ground state energy of $H^{\prime}.$ \ Since
$H^{\prime}$ and $\frac{1}{\left|  G\right|  }H$ are equivalent within the
$G$-invariant subspace, we conclude from (\ref{bound}) that the energy
spectrum of $H$ is bounded below by $\left|  G\right|  \cdot E_{0}^{\prime}$.

\section{Numerical example}

We\ now numerically check the lower energy bound for our one-dimensional
chains. \ We use the notation $E_{0}^{N}(c_{1},c_{2})$ for the ground state
energy of the periodic chain Hamiltonian in (\ref{hamilton}).$\ \ $This is
trivially calculated in our example by momentum decomposition,%
\begin{equation}
E_{0}^{N}(c_{1},c_{2})=\sum_{k=0,...,N-1}\min\left(  2c_{1}\cos(\tfrac{2k\pi
}{N})+c_{2},0\right)  .
\end{equation}
\ It easy to see that the lowest point of the spectrum in the translationally
invariant subspace is also $E_{0}^{N}(c_{1},c_{2})$. \ We take $E_{0}^{\prime
M}(\tfrac{c_{1}}{M-1},\tfrac{c_{2}}{M})$ to be the ground state energy of the
non-periodic chain Hamiltonian of length $M$. \ We have renormalized the
coefficients as prescribed in (\ref{renorm}). \ In Tables 1 and 2 we show
$E_{0}^{N}(c_{1},c_{2})$ and the lower energy bound $N\cdot E_{0}^{\prime
M}(\tfrac{c_{1}}{M-1},\tfrac{c_{2}}{M})$ for several values of $N$, $M$,
$c_{1}$, and $c_{2}$. \ The values for $E_{0}^{\prime M}(\tfrac{c_{1}}%
{M-1},\tfrac{c_{2}}{M})$ were calculated by Lanczos diagonalization.%
\[%
\begin{array}
[c]{c}%
\text{Table 1. \ Lower bounds and energies for }c_{1}=-1,c_{2}=0\\%
\begin{tabular}
[c]{|l|l|l|l|l|l|}\hline
$N$ & $N\cdot E_{0}^{\prime4}$ & $N\cdot E_{0}^{\prime8}$ & $N\cdot
E_{0}^{\prime12}$ & $N\cdot E_{0}^{\prime16}$ & $E_{0}^{N}$\\\hline
$16$ & $-11.93$ & $-10.88$ & $-10.61$ & $-10.49$ & $-10.05$\\\hline
$32$ & $-23.85$ & $-21.75$ & $-21.23$ & $-20.99$ & $-20.31$\\\hline
$64$ & $-47.70$ & $-43.51$ & $-42.45$ & $-41.98$ & $-40.71$\\\hline
$128$ & $-95.41$ & $-87.02$ & $-84.90$ & $-83.95$ & $-81.47$\\\hline
\end{tabular}
\end{array}
\]%
\[%
\begin{array}
[c]{c}%
\text{Table 2. \ Lower bounds and energies for }c_{1}=-1,c_{2}=1\\%
\begin{tabular}
[c]{|l|l|l|l|l|l|}\hline
$N$ & $N\cdot E_{0}^{\prime4}$ & $N\cdot E_{0}^{\prime8}$ & $N\cdot
E_{0}^{\prime12}$ & $N\cdot E_{0}^{\prime16}$ & $E_{0}^{N}$\\\hline
$16$ & $-4.63$ & $-4.08$ & $-3.90$ & $-3.76$ & $-3.53$\\\hline
$32$ & $-9.26$ & $-8.17$ & $-7.79$ & $-7.52$ & $-7.00$\\\hline
$64$ & $-18.52$ & $-16.33$ & $-15.59$ & $-15.05$ & $-13.96$\\\hline
$128$ & $-37.04$ & $-32.67$ & $-31.18$ & $-30.10$ & $-27.91$\\\hline
\end{tabular}
\end{array}
\]
We note that the error between the lower bound estimate and the actual energy
scales as $M^{-1}$, the ratio of the boundary size to the bulk size of the
cluster. \ This is also expected in the general case.

\section{Summary}

We have derived lower energy bounds for any translationally invariant lattice
Hamiltonian. \ Our result is general and applies to virtually any local
quantum lattice Hamiltonian relevant to particle, nuclear, or many-body
physics. \ The lower bound is given in terms of the ground state energy for a
renormalized Hamiltonian defined on a subset of lattice points. \ We find that
the error scales as the ratio of the boundary size to the bulk size of the
cluster. \ Given the simplicity, generality, and convenience of the result, we
hope this lower bound estimate will be useful in many future computational
lattice calculations.


\begin{thebibliography}{9}
\bibitem{lee}D. J. Lee, N. Salwen, D. D. Lee, Phys. Lett. B503 (2001) 223; D.
Lee, N. Salwen, M. Windoloski, Phys. Lett. B502 (2001) 329.

\bibitem{white}S. White, Phys.\ Rev. Lett. 69 (1992) 2863; Phys. Rev. B48
(1993) 10345.

\bibitem{hirsch}J. Hirsch, R. Sugar, D. Scalapino, R. Blankenbecler, Phys.
Rev.\ Lett. 46, (1981) 1628; Phys. Rev. B26 (1982) 5003.

\bibitem{anderson}P.W. Anderson, Phys. Rev. 83, 1260 (1951).

\bibitem{wittman}T. Wittmann, J. Stolze, Phys. Rev. B 48 3479 (1993).

\bibitem{valenti}R. Tarrach, R. Valenti, Phys. Rev. B 41, 9611 (1990).
\end{thebibliography}
\end{document}